\documentclass{elsart}
\usepackage{graphicx,amssymb}
\journal{Physics Letters B}

\def\lbar{\mathrel{\rlap{
\raise3pt\hbox{\hskip-1.75pt-}}
\hbox{$\lambda_e$}}}
\begin{document}
\begin{frontmatter}

\title{Effects of Strong Magnetic Fields in Strange Baryonic Matter}
\author[Caltech]{A.E. Broderick}
\ead{aeb@tapir.caltech.edu}
\author[SUNYSB]{M. Prakash}
\ead{prakash@snare.physics.sunysb.edu}
\author[SUNYSB]{and J.M. Lattimer}
\ead{lattimer@astro.sunysb.edu}

\address[Caltech]{Theoretical Astrophysics, 
California Institute of Technology, 
1200 East California Blvd, Pasadena, CA 91125}
\address[SUNYSB]{Department of Physics and Astronomy, 
State University of New York at Stony Brook, 
Stony Brook, NY 11974-3800}

\begin{abstract}
We investigate the effects of very strong magnetic fields upon the
equation of state of dense bayonic matter in which hyperons are
present.  In the presence of a magnetic field, the equation of state
above nuclear density is significantly affected both by Landau
quantization and magnetic moment interactions, but only for field
strengths $B>5\times10^{18}$ G.  The former tends to soften the EOS
and increase proton and lepton abundances, while the latter produces
an overall stiffening of the EOS.  Each results in a supression of
hyperons relative to the field-free case.  The structure of a neutron
star is, however, primarily determined by the magnetic field stress.
We utilize existing general relativistic magneto-hydrostatic
calculations to demonstrate that maximum average fields within a
stable neutron are limited to values $B\le 1-3 \times10^{18}$ G.  This
is not large enough to significantly influence particle compositions
or the matter pressure, unless fluctuations dominate the average field
strengths in the interior or configurations with significantly larger
field gradients are considered.
\end{abstract} 
\end{frontmatter}
\noindent PACS numbers: 26.60+c, 97.60.Jd, 98.35.Eg   \newpage

Recent discoveries that link soft $\gamma$-ray repeaters and perhaps
some anomalous X-ray pulsars with neutron stars having ultrastrong
magnetic fields - the so called magnetars (see Table 1 in \cite{CPL01}
for a summary) - have spurred theoretical studies of the effects
ultrastrong magnetic fields have on the equation of state (EOS) of
neutron-star matter and on the structure of neutron stars
(cf. \cite{CPL01,Bocq95,BPL00} and references therein).  The magnetic
field strength $B$ needed to dramatically affect neutron star
structure directly can be estimated with a dimensional analysis
\cite{LS91} equating the magnetic field energy $ E_B \sim (4\pi
R^3/3)(B^2/ 8\pi) $ with the gravitational binding energy $ E_{B.E.}
\sim 3GM^2/5R$, yielding
\begin{eqnarray}
B \sim 1.4\times
10^{18}\left(M\over 1.4~{\rm M}_\odot\right) \left(R\over 10\ 
{\rm km}\right)^{-2}{\rm~G} \,,
\label{fest}
\end{eqnarray}
where $M$ and $R$ are, respectively, the neutron star mass and radius.

To date, studies of the effects of such ultrastrong magnetic fields on
the EOS and on the structure of neutron stars have been largely
limited to cases in which the strongly interacting component of matter
is comprised of nucleons~\cite{BPL00,Cha96,CBP97,YZ99}.  Our objective
in this paper is to investigate the effects of magnetic fields on the
EOS of matter containing strangeness-bearing hyperons.  Towards this
end, we utilize the theoretical formalism for the EOS developed in
Ref.~\cite{BPL00}.  This allows us to consistently incorporate the
effects of magnetic fields on the EOS in multicomponent, interacting
matter.  For the first time, the anomalous magnetic moments of baryons
is included, using a covariant description.  A further objective is to
explore the consequences of strong magnetic fields for neutron star
structure and maximum masses, which is accomplished using the results
of general relativistic structural calulations in Ref.~\cite{CPL01}.

Charge neutral, beta-equilibrated, neutron-star matter contains both
negatively charged leptons ($e$ and $\mu$) and positively charged
baryons ($p$ and, at higher densities, $\Sigma^+$).  Magnetic fields
quantize the orbital motion (Landau quantization) of these charged
particles.  When the Fermi energy of the $p$ or $\Sigma^+$ becomes
significantly altered by the magnetic field, the pressure and
composition of matter in beta equilibrum are affected.  Landau
quantzation has a significant effect when $B^*=B/B^e_c \sim 10^5$
($B_c^e= \hbar c/(e\lbar^2) = 4.414\times10^{13}$ G is the critical
electron field)~\cite{BPL00,Cha96,CBP97,YZ99}.  Higher fields lead to
a general softening of the EOS relative to the case in which magnetic
fields are absent.

In strong magnetic fields, contributions from the anomalous magnetic
moments of the baryons (denoted hereafter by $\kappa_b$) must also be
considered.  The measured magnetic moments $\mu_b$ for nucleons and
strangeness-bearing hyperons are given in Table 1.  With few
exceptions, the anomalous magnetic moments are similar in magnitude.
As discussed below, the energies $|\kappa_n B|\simeq 2.7\times10^{-4}
B^*$ MeV and $|\kappa_p B|\simeq 2.5\times10^{-4} B^*$ MeV measure the
changes to the nucleon Fermi energies, and $|\kappa_n + \kappa_p| B
\simeq 1.67 \times 10^{-5} B^*$ MeV measures the change to the beta
equilibrium condition.  Since the Fermi energies exceed 20 MeV for the
densities of interest, it is clear that contributions from the
anomalous magnetic moments become significant for $B^* > 10^5$.  For
such large fields, complete spin polarization of the neutrons occurs,
producing an increase in degeneracy pressure and an overall stiffening
of the EOS that overwhelms the softening induced by Landau
quantization~ \cite{BPL00}.

Similary, in the presence of strong magnetic fields, all of the
hyperons are susceptible to spin polarization due to magnetic moment
interactions with the field.  The net result of the opposing effects
of degeneracy and Landau quantization will depend on the relative
concentrations of the various hyperons, which in turn depend
sensitively on the hyperon-meson interactions for which only a modest
amount of guidance is available~\cite{GM91,KPE95,SM96}.  It is one of
the purposes of this work to ascertain the influence of feedback
effects due to mass and energy shifts in multi-component matter.

\begin{table}[!h]
\begin{center}
\noindent{TABLE 1: The measured magnetic moments of spin$-\frac 12$ baryons 
from Ref.~\cite{Data}.} 
$\mu_N=e\hbar/(2m_p)=3.15\times10^{-18}$ MeV
G$^{-1}$ is the nuclear magneton and the anomalous magnetic moment 
$\kappa_b = (\mu_b/\mu_N - 
q_b m_p/m_b)~\mu_N$. \\ 
\begin{tabular}{llrrr}
\hline
Species & Mass & Charge &  Magnetic Moment & 
Anomalous Moment \vspace*{-.1in}
\cr
$b$ & (MeV) & $q_b$ & $\mu_b/\mu_N$ & $\kappa_b / \mu_N$ \\
\hline
$p$ & $938.3$ &  $1$ & $2.79$ & $1.79$ \\ 
$n$ & $939.6$ &  $0$ & $-1.91$ & $-1.91$ \\
\hline
$\Lambda$ & $1115.7$  & $0$ & $-0.61$ & $-0.61$ \\
\hline
$\Sigma^+$ & $1189.4$ &  $1$ & $2.46$ & $1.67$ \\
$\Sigma^0$ & $1192.6$ &  $0$ & $1.61$ & $1.61 $ \\
$\Sigma^-$ & $1197.4$ &  $-1$ & $-1.16$ & $-0.38 $ \\
\hline
$\Xi^0$ & $1314.9$ &  $0$ & $-1.25$ & $-1.25$ \\
$\Xi^-$ & $1321.3$ &  $-1$ & $-0.65$ & $0.06$ \\
\hline
\end{tabular}
\label{moments}
\end{center}
\end{table}

To describe the EOS of neutron-star matter, we employ a field
theoretical approach (at the mean field level) in which the baryons
interact via the exchange of $\sigma$-$\omega$-$\rho$
mesons.\footnote{ The qualitative effects of strong magnetic fields
found in the field-theoretical models also exist in non-relativistic
potential models. This is because the phase space of charged particles
is similarly affected in both approaches by magnetic fields.  The
effects due to the anomalous magnetic moments would, however, enter
linearly in a non-relativistic approach, and would thus be more
dramatic in that case.}  We consider two classes of models that differ
in their high density behavior.  The two Lagrangian densities are
given by \cite{BB77,ZM90}
\begin{eqnarray}
{\mathcal L}_{I} &=& {\mathcal L}_b -
\left( 1 - \frac{g_{\sigma b} \sigma}{m_b}
\right) \overline{\Psi}_b m_b \Psi_b  + {\mathcal L}_m + {\mathcal L}_\ell 
\,,\nonumber\\
{\mathcal L}_{II} &=& \left( 1 + \frac{g_{\sigma b} \sigma}{m_b} \right) 
{\mathcal L}_b
-  \overline{\Psi}_b m_b \Psi_b  + {\mathcal L}_m + {\mathcal L}_\ell \,.
\end{eqnarray}
The baryon, lepton, and meson Lagrangians are given by
\begin{eqnarray}
{\mathcal L}_b &=& \overline{\Psi}_b \left( i \gamma_\mu \partial^\mu
+q_b \gamma_\mu A^\mu - g_{\omega b} \gamma_\mu \omega^{\mu}
- g_{\rho b} \tau_{3_b} \gamma_\mu \rho^{\mu}
- \kappa_b \sigma_{\mu \nu} F^{\mu \nu}
\right) \Psi_b \,,\nonumber \\
{\mathcal L}_\ell &=& \overline{\psi}_l \left( i \gamma_\mu \partial^\mu
+q_l \gamma_\mu A^\mu  \right) \psi_l \,, \quad {\rm and} \nonumber \\
{\mathcal L}_m &=& \frac{1}{2} \partial_{\mu} \sigma \partial^{\mu} \sigma
- \frac{1}{2} m_{\sigma}^2 \sigma^2 - U(\sigma)
 + \frac{1}{2} m_{\omega}^2 \omega_{\mu} \omega^{\mu}
  - \frac{1}{4} \Omega^{\mu \nu} \Omega_{\mu \nu} \nonumber \\
 & & \mbox{} + \frac{1}{2} m_{\rho}^2 \rho_{\mu} \rho^{\mu}
- \frac{1}{4} P^{\mu \nu} P_{\mu \nu}  -\frac{1}{4} F^{\mu \nu} F_{\mu \nu} \,,
\label{Lag}
\end{eqnarray}
where $\Psi_b$ and $\psi_\ell$ are the baryon and lepton Dirac fields,
respectively.  The index $b$ runs over the baryons $n$, $p$,
$\Lambda$, $\Sigma^-$, $\Sigma^0$, $\Sigma^+$, $\Xi^-$, and
$\Xi^0$. (Neglect of the $\Omega^-$ and the $\Delta$ quartet, which
appear only at very high densities, does not qualitatively affect our
conclusions.)  The index $\ell$ runs over the electron and muon.  The
baryon mass and the isospin projection are denoted by $m_b$ and
$\tau_{3_b}$, respectively.  The mesonic and electromagnetic field
tensors are given by their usual expressions: $\Omega_{\mu\nu} =
\partial_\mu \omega_\nu - \partial_\nu \omega_\mu$, $P_{\mu\nu} =
\partial_\mu \rho_\nu - \partial_\nu \rho_\mu$, and $F_{\mu\nu} =
\partial_\mu A_\nu - \partial_\nu A_\mu$.  The strong interaction
couplings are denoted by $g$, the electromagnetic couplings by $q$,
and the meson masses by $m$ all with appropriate subscripts.  The
quantity $U(\sigma)$ denotes scalar self-interactions and is taken to
be of the form $ U(\sigma) = (b m_n/3) (g_{\sigma N} \sigma)^3 + (c/4)
(g_{\sigma N} \sigma)^4$~\cite{BB77,Gl82,Gl85}.

The anomalous magnetic moments are introduced via the minimal coupling
of the baryons to the electromagnetic field tensor for each
baryon.  Their contributions in the Lagrangian are contained in the
term $ - \kappa_b \overline{\Psi}_b \sigma_{\mu \nu} F^{\mu \nu}
\Psi_b$, where $\sigma_{\mu \nu} = \frac{i}{2}
\left[\gamma_{\mu},\gamma_{\nu} \right]$.  Although the
electromagnetic field is included in ${\mathcal L}_I$ and ${\mathcal
L}_{II}$, it assumed to be externally generated and only frozen-field
configurations will be considered.

The meson-nucleon $g_{iN} (i=\sigma,\omega,\rho)$, and scalar
self-interaction couplings, $b$ and $c$, were chosen to reproduce the
binding energy $B/A$, the nuclear saturation density $n_s$, the Dirac
effective mass $M^*$, and the symmetry energy $a_s$.  The values of
the couplings used in the models considered are shown in Table 2, and
the associated nuclear matter properties used to calculate these
couplings are shown in Table 3.

The meson-hyperon couplings are assumed to be fixed fractions of the
meson-nucleon couplings, i.e., $g_{iH} = x_{iH} g_{iN}$, where for
each meson $i$, the values of $x_{iH}$ are assumed equal for all
hyperons $H$.  The values of $x_{iH}$ are chosen to reproduce the
binding energy of the $\Lambda$ at nuclear saturation as suggested by
Glendenning and Moszkowski (GM)~\cite{GM91} and are also given in
Table 2.  The high-density behavior of the EOS is sensitive to the
strength of the meson couplings employed~\cite{KPE95} and the models
chosen encompass a fairly wide range of variation.  Models GM1--3
employ linear scalar couplings (${\mathcal L}_I$), while the Zimanyi
and Moszkowski (ZM) model employs a nonlinear scalar coupling
(${\mathcal L}_{II}$), which is reflected in the high density
behaviors of $m_n^*/m_n$.  Comparison of results from the GM1--3 and
ZM models allows us to contrast the effects of the underlying EOS.

\begin{table}
\begin{center}
\noindent{TABLE 2: Nucleon-meson and hyperon-meson coupling constants
for the GM1--3 and ZM models.  The
hyperon to nucleon coupling ratios were taken from 
Ref.~\cite{PCL95}.}\\
\begin{tabular}{cccccccccr}
\hline
Model & $g_{\sigma N}/m_\sigma $ & $g_{\omega N}/m_\omega$ 
& $g_{\rho N}/m_\rho$ &  $x_{\sigma H}$& $x_{\omega H}$ 
& $x_{\rho H}$ & $b$ &  $c$ \vspace*{-.1in}\cr
 & (fm) & (fm) & (fm) & & &  & & \\
\hline
 GM1 & 3.434 & 2.674 & 2.100 & 0.6 & 0.653 & 0.6 & 0.002947 &
 $-0.001070$ \\
 GM2 & 3.025 & 2.195 & 2.189 & 0.6 & 0.659 & 0.6 & 0.003478 &
 0.01328 \\
 GM3 & 3.151 & 2.195 & 2.189 & 0.6 & 0.659 & 0.6 & 0.008659 &
 $-0.002421$ \\
\hline
 ZM  & 2.736 & 1.617 & 2.185 & 1.0 & 1.0 & 1.0 & 0.0 & 0.0 \\
\hline
\hline
\end{tabular}
\label{couplings}
\end{center}
\end{table}

\begin{table}
\begin{center}
\noindent{TABLE 3: Nuclear matter properties used to fit 
coupling constants for the GM1--3 Glendenning and Moszkowski (GM1-3)   
Zimanyi and Moszkowski (ZM) models.} \\
\begin{tabular}{ccccccr}
\hline
Model & $n_s$ & $-B/A$ & $M^*/M$ & $K_0$ & $a_{\mbox{\tiny sym}}$  
\vspace*{-.1in}\cr
 & (fm$^{-3})$ & (MeV) & & (MeV) & (MeV) \\
\hline
 GM1 & 0.153 & 16.30 & 0.70 & 300 & 32.5 \\
 GM2 & 0.153 & 16.30 & 0.78 & 300 & 32.5 \\
 GM3 & 0.153 & 16.30 & 0.78 & 240 & 32.5 \\
\hline
 ZM  & 0.160 & 16.00 & 0.86 & 225 & 32.5 \\
\hline
\hline
\end{tabular}
\label{model_props}
\end{center}
\end{table}

The field equations (not explicitly displayed here) follow from a
natural extension of those used in Ref.~\cite{BPL00} to the case in
which hyperons are present.  In charge neutral and beta equilibrated
matter, the conditions
\begin{equation}
\sum_b q_b n_b + \sum_l q_l n_l = 0 \qquad {\rm and} \qquad 
\mu_b = B_b \mu_n - q_b \mu_e \,,
\label{equil}
\end{equation}
where $\mu_b$ is the baryon chemical potential (not to be confused
with the magnetic moment $\mu_b$ of Table~1) and $B_b$ is the baryon
number, also apply.  The energy spectra, number densities, scalar
number densities, energy densities, and total pressure are given by
straightforward generalizations of their analogues in
Ref. \cite{BPL00}.

\begin{figure}[!ht]
\begin{center}
\includegraphics*[width=15cm]{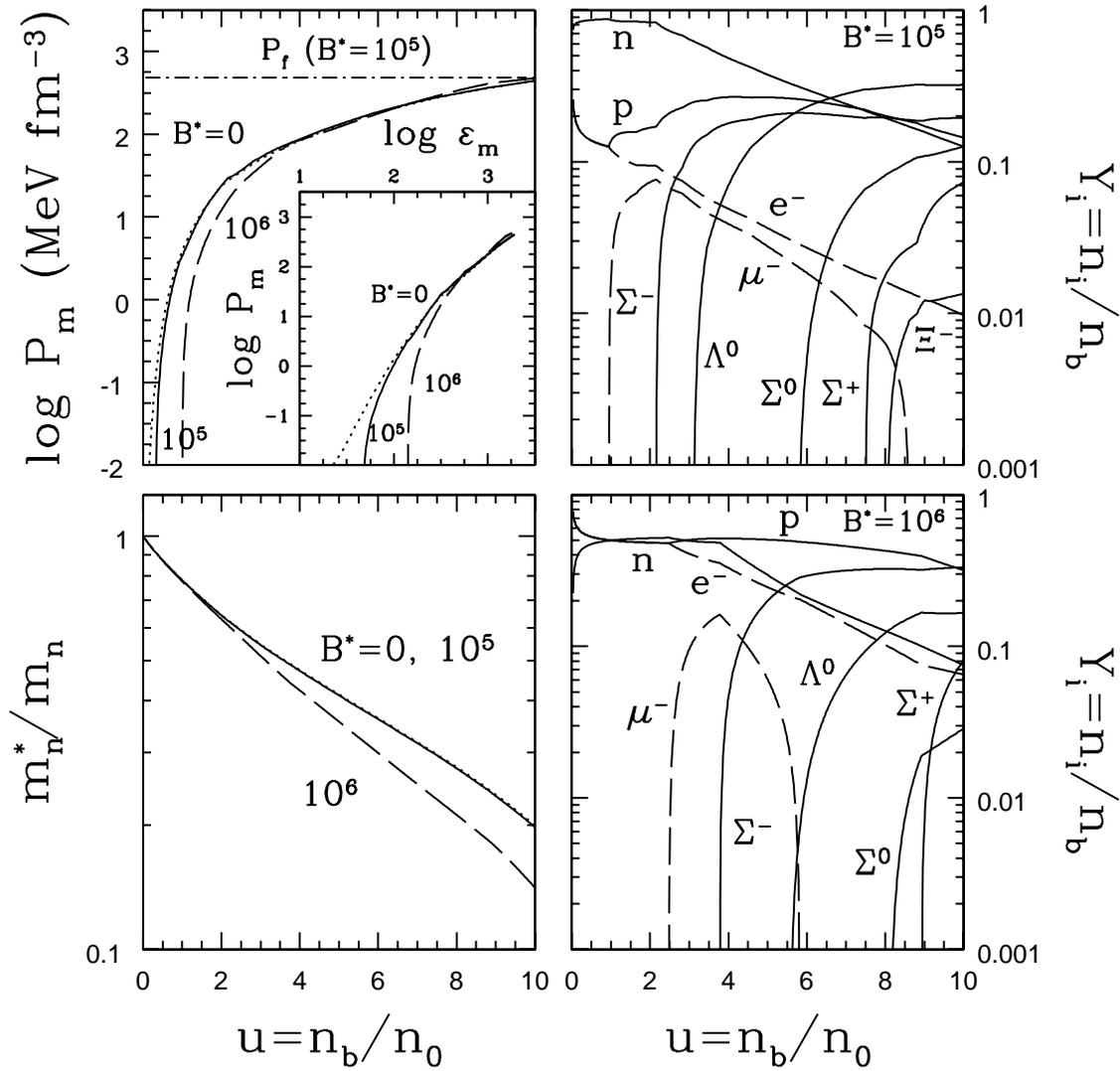}
\end{center}
\caption{Matter pressure $P_m$, neutron Dirac effective mass
$m^*_n/m_n$, and concentrations $Y_i$ as functions of density ($n_0 =
0.16$ fm$^{-3}$ is the fiducial nuclear saturation density) for the
model GM3.  Each is shown for magnetic field strengths of $B^*=0$,
$10^5$, and $10^6$.  
$P_f=B^2/8\pi$ 
is the pressure due to magnetic
field stress for the value $B^*=10^5$.  The inset shows $P_m$ as a
function of the matter energy density $\varepsilon_m$.  In the right
panels results for leptons are shown by dashed lines while those for
baryons are shown by solid lines.}
\label{GM3_LL}
\end{figure}

In Fig.~\ref{GM3_LL}, results are shown for physical quantities of
interest for the baseline model GM3, when the effects of Landau
quantization are included but those of anomalous magnetic moments are
ignored.  As the magnetic field is varied, 
the generic behavior of the matter pressure $P_m$, effective neutron mass
$m^*_n$, and particle abundances are the same as those found  in
Ref.~\cite{BPL00}, in which matter containing only nucleons was
considered.  When hyperons appear, the pressure becomes smaller than
the case of pure nucleons for all fields.  The effective nucleon mass
decreases slightly when hyperons appear, again independently of the
field strength.  The appearance of hyperons produces pronounced
effects on the nucleon and lepton abundances.  

The cases $B^*=0$ and $B^*=10^5$ are nearly indistinguishable, except
for densities less than nuclear density ($u<1)$, where ``cavitation''
due to Landau quantization occurs.  This effect has been discussed in
Ref.~\cite{BPL00} and earlier work (cf., Refs. \cite{Cha96,CBP97}).
However, the structure of a magnetized neutron star will be mostly
affected by contributions from the magnetic field stress,
$P_f=B^2/8\pi = 4.814\times 10^{-8}B^{*^2}~{\rm MeV~fm}^{-3}$, which
greatly exceeds the matter pressure $P_m$ at all relevant densities
for $B^*\ge 10^5$, as shown in Fig. 1 for $B^*=10^5$.  

Hyperons appear near twice nuclear density for $B^*\le10^5$ in this
model.  For $B^*>10^5$ ($B>5\times10^{18}$ G), the same value 
needed to noticeably affect the matter pressure and nucleon abundances,
significant effects on hyperon threshold densities and abundances are
apparent.  Landau quantization increases the proton abundance, and
hence the electron concentration due to charge neutrality.  Therefore
$\mu_e=\mu_n-\mu_p$ is increased.  However, at fixed baryon density,
the neutron abundance must then decrease, which significantly lowers
$\mu_n$ to a greater extent than $\mu_n-\mu_p$ is increased.  The net
effect is a suppression of hyperons.  In particular, note that
$\mu_{\Sigma^-}=2\mu_n-\mu_p$ and $\mu_{\Lambda}=\mu_n$.

It should also be noted that, compared to the case without hyperons
(cf. Ref.~\cite{BPL00}), the pressure is reduced for all field
strengths, as is the nucleon effective mass.  The magnitude of the
nucleon effective mass can be important in determining the chemical
potentials, as shown below.

\begin{figure}[!h]
\begin{center}
\includegraphics*[width=15cm]{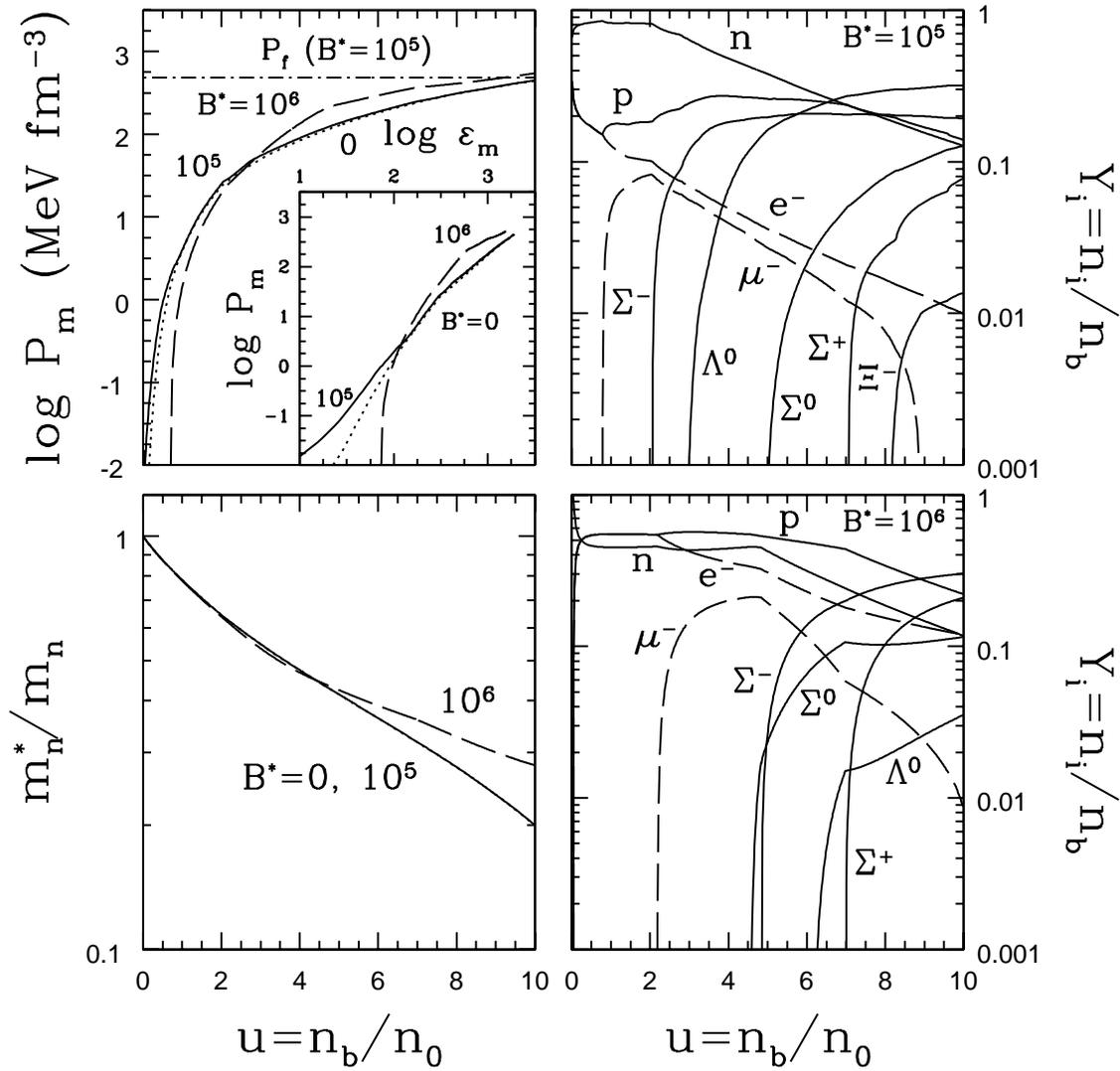}
\end{center}
\caption{Same as Fig.~\ref{GM3_LL}, but with magnetic moment
interactions included.}
\label{GM3_MM}
\end{figure}

We have verified that the qualitative effects of large fields on the
composition, presure and effective mass are similar for the other
three models considered, in the case that only Landau
quantization is considered. The relative suppression of individual
hyperons varies, however; for example in models GM2 and ZM the
$\Lambda$ and $\Sigma^-$ threshold densities reverse at large fields
relative to GM1 and GM3.  Importantly, however, the value $B^*\sim
10^5$ represents a threshold for field effects to become significant,
independently of the EOS.

The inclusion of anomalous magnetic moments produces even larger
suppressions of hyperons, as displayed in Fig.~\ref{GM3_MM} for the
baseline model GM3.  One can qualitatively understand this trend by
considering the effects of magnetic moments on the baryon chemical
potentials, and then examining the effects on the threshold of the
first hyperon to appear.  At low temperatures, the chemical potentials
can be identified with the Fermi levels of the baryonic energy
spectra.  The energy spectrum for charged baryons is given
by~\cite{BPL00}
\begin{eqnarray}
E_{b,n,s} &=& \sqrt{ k_z^2 + \left( \sqrt{m_b^{*~2} + 2 \nu q_b B}
+ s \kappa_p B \right)^2 }
 + g_{\omega_p} \omega^0 + t_{3_b} g_{\rho_p} \rho^0 \,, \nonumber \\
&\simeq &  m_b^* + \frac {k_z^2}{2 m_b^*}
+  \frac{\nu q_b B}{m_b^*} + s \kappa_b B +  
g_{\omega_b} \omega^0  +t_{3_b}g_{\rho_b} \rho^0 \,,
\label{ech_kap}
\end{eqnarray}
where the quantity $\nu = n+s/2+1/2$ characterizes the so-called
Landau level with $n$ and $s~(\pm 1)$ being the principle quantum
number and ``spin'' quantum number respectively, and $t_{3_b}$ is the
isospin projection of baryon $b$. The quantities $\omega^0$ and
$\rho^0$ are mean field solutions of the field equations corresponding
to Eq.~(\ref{Lag})~(cf. Ref.~\cite{BPL00}).  The rightmost expression
is valid in the nonrelativistic limits $k_z^2 \ll m_b^{*^2}$, $2\nu
|q_b|B \ll m_b^{*^2}$ and $(\kappa_b B)^2 \ll m_b^{*^2}$, which are
appropriate for the conditions of interest.  Similarly, the energy
spectrum of neutral baryons is given by
\begin{eqnarray}
E_{b,s} &=& \sqrt{ k_z^2 + \left( \sqrt{m_b^{*~2} + k_x^2 + k_y^2} +
s \kappa_b B \right)^2}  + g_{\omega_b} \omega^0 + 
t_{3_b} g_{\rho_b} \rho^0\,, \nonumber \\
&\simeq & m_b^* + \frac {k^2}{2 m_b^*}
+ s \kappa_b B +  
g_{\omega_b} \omega^0 + t_{3_b} g_{\rho_b} \rho^0\,.
\label{en_kap}
\end{eqnarray}

For both charged and neutral baryons, the chemical potentials of
nearly degenerate systems are set by the lowest energy spin state.
The contribution to neutral baryon energies, and to their chemical
potentials, due to the anomalous magnetic moment is therefore
$-|\kappa_b| B$ in Eq.~(\ref{en_kap}).  The situation of charged
baryons is more complicated, because of the additional contribution
due to the Dirac moment.  However, given that the value of $n$ from
Landau quantization generally exceeds $|\kappa_b|/\mu_b$, the shift in
charged baryon chemical potentials is not as significant as it is for
neutral baryons.  In matter containing only nucleons, the predominant
effect of the anomalous magnetic moments is therefore to decrease
$\mu_n$, and thus the neutron abundance, relative to the case in which
the anomalous moments are ignored (compare Figs.~\ref{GM3_LL} and
\ref{GM3_MM}).  A further consequence is that the threshold 
density for the first hyperon, $\Sigma^-$, is increased since
$\mu_{\Sigma^-}=2\mu_n-\mu_p$.  Although the chemical potential of the
$\Sigma^-$ also receives a contribution from its anomalous magnetic
moment, the magnitude of the change is small compared to that due to
the neutron.  Note again that field strengths $B^*>10^5$
are necessary to produce appreciable effects.  We have verified that
these conclusions hold for the other three EOSs we considered for the
case when magnetic moments are included.

While most hyperon threshold densities are increased by the inclusion
of anomalous magnetic moments, those of $\Sigma^0$ and $\Sigma^+$, the
latter to a lesser extent, are decreased for $B^*>10^5$.  The complex
interplay that exists when several baryons are simultaneously present
makes a simple explanation difficult, however.

Finally, as noted in Ref. ~\cite{BPL00}, the inclusion of magnetic
moments produces an increase in pressure and effective mass at large
fields ($B^*>10^5$), due to increased baryon degeneracies
~\cite{BPL00}.  These increases more than offset the reductions
induced by the inclusion of hyperons. 

\begin{figure}[!h]
\begin{center}
\includegraphics*[width=12cm]{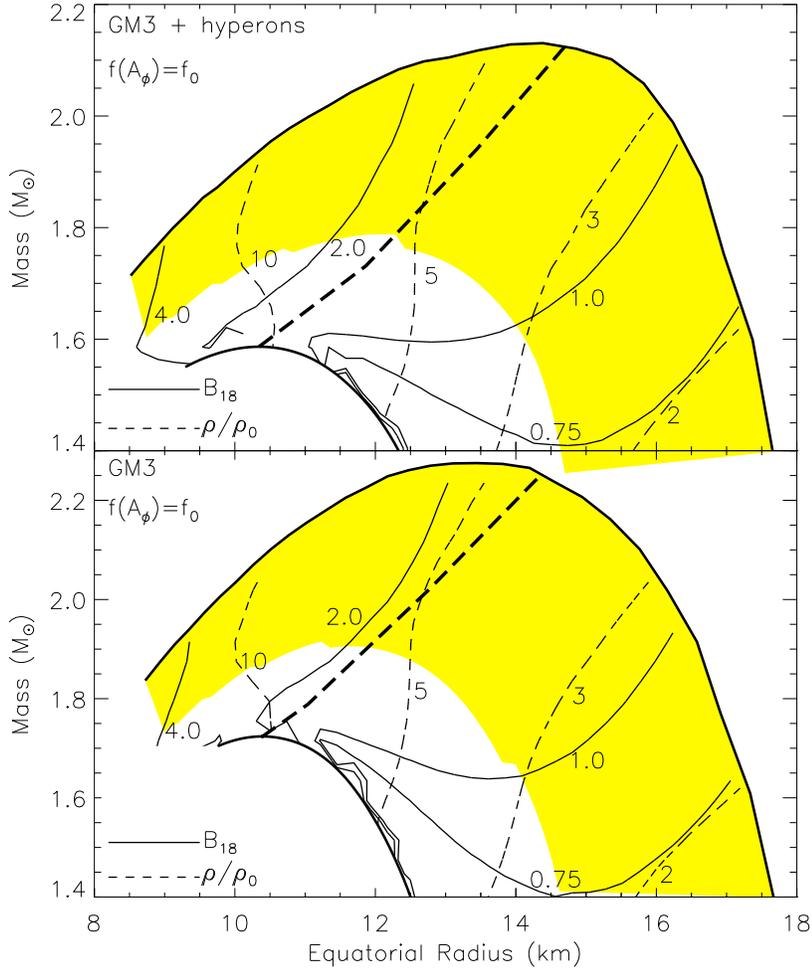}
\end{center}
\caption{Limits to hydrostatic configurations for neutron stars permeated by
axially symmetric magnetic fields.  The upper (lower) panel is for the
EOS GM3 including (excluding) hyperons.  In each panel,
the lower heavy solid curve is the standard mass-radius relation for
field-free stars.  The upper heavy solid curve represents the largest
gravitational mass possible for a given equatorial radius as the
magnetic fields are increased, for the indicated current function
$f$.  The heavy dashed curve is the locus of minima of the contours of
fixed baryon mass configurations: Ref.~\cite{CPL01} suggested that
this could be the limit to dynamical stability. Thin solid lines are
contours of the maximum magnetic field strength in units of $10^{18}$
G.  Similarly, thin dashed lines are contours of
maximum mass-energy density in units of the energy density $\rho_0$ at the
nuclear saturation density.}
\label{jimc}
\end{figure}

Still to be addressed, however, is whether or not stable stellar
configurations can exist in which the magnetic field is large enough
($B>5\times10^{18}$ G) that the properties of matter are significantly
affected by the magnetic field.  One step in this direction was
recently undertaken in Ref.~\cite{CPL01}, in which the limits of
hydrostatic equilibrium for axially-symmetric magnetic fields in
general relativistic configurations were analyzed.  As discussed in
detail in Ref. \cite{Bocq95}, in axially symmetric field
configurations with a constant current function, the magnetic field
contributes a centrifugal-like contribution to the total stress
tensor.  This can be understood by noting that for the artificial
field geometry considered (namely $\vec{B} \propto \hat{z}$), a
superconducting fluid can move along field lines but not across them.
Thus the ``pressure'' associated with the magnetic field will only act
equatorially and not vertically.  This flattens an otherwise spherical
star, and for large enough fields, decreases the central (energy)
density even as the mass is increased.  For large enough fields, the
star's stability is eventually compromised.

Hydrostatically stable configurations (of which some may not be 
stable to dynamical perturbations) for the GM3 EOS, both excluding and
including hyperons, are shown in mass-equatorial radius space in
Fig.~\ref{jimc}.  The lower thick black line in each panel is the
usual field-free, spherical result for the mass-radius relation.  The
upper thick black line represents the largest possible stable mass for
a given equatorial radius as the internal field is increased.  Large
axially-symmetric fields tend to yield flattened configurations, and
if large enough, shift the maximum densities off-center in the shaded
regions in the figure.  This results in toroidal shapes with low-density
centers.  As discussed in Ref.~\cite{CPL01}, regions to the left of
the thick dashed line in this figure are likely unstable to small
amplitude perturbations and hence unrealistic as stable physical
configurations.  

Superimposed on this figure are contours of maximum
mass-energy density and magnetic field strength.
For the GM3 model, hyperons appear at zero field at about twice
nuclear density.  Therefore, the portions of the two panels in which
the maximum density is below about 3 times nuclear density are nearly
identical.  For the particular choice of a constant current
function, but independently of the EOS, it is found that the maximum
value of the magnetic field within the star cannot exceed about
$3\times10^{18}$ G, which corresponds to $B^*\sim 7\times10^4$. In the case
of GM3, the maximum value is $B\sim1.8\times10^{18}$ G.  This
field strength is not nearly large enough to produce appreciable
effects on hyperon or nucleon compositions, with or without the
inclusion of anomalous magnetic moments.  The dominant effect of the
field arises through the magnetic field stress, which effectively
dominates the matter pressure below a few times nuclear saturation
density, depending on the field's orientation.  
Whether other choices of
current functions, or the relaxation of the condition of axial
symmetry, 
will alter these conclusions is not clear, as the answer will likely
depend on the detailed nature of the field configurations considered.

To date only a few cases of spatially varying current functions have
been explored~\cite{Bocq95}, but these have all been limited to axial
symmetry.  Furthermore, the cases explored have relatively small
spatial gradients so that the ratio of the maximum field to the
average field within the star is not large.  It might be that the
shapes of the stars will significantly change with a different field
geometry.  It is even possible to imagine a disordered field where
$\langle B^2\rangle$ is significantly larger than 
$\langle \vec{B} \rangle^2$.  
In this case the pressure will be dominated by fluctuations in the field, 
but the stars will tend to be spherical.  It is possible that strong 
magnetic fields may be held in the core for periods much longer than the 
ohmic diffusion time due to interactions between the magnetic flux tubes
and the vortex tubes expected to be present in a superconducting, 
superfluid, rotating neutron star~\cite{Rud95}.
Therefore, although the results to date imply that average
fields within a neutron star cannot exceed $1-3\times10^{18}$ G before
stability is compromised, the value of the maximum possible field at
any point within the star is undetermined.  As a result, it is possible
that the effects of magnetic fields upon the equation of state or particle
composition will be important in some magnetic field geometries.

Research support from DOE grants FG02-88ER-40388 (for MP) and
FG02-87ER40317 (for JML) are gratefully acknowledged. We are grateful
to Christian Cardall for providing us numerical data for the construction
of Fig.~\ref{jimc}.  We would like to thank Yasser Rathore and Greg 
Ushomirsky for helpful discussions.

\end{document}